\documentclass[11pt,a4paper]{article}
\usepackage{jheppub}

\usepackage{slashed}
\usepackage{hyperref}
\usepackage{epsfig,amsmath,amsfonts,amssymb,amsthm}
\usepackage[normalem]{ulem}
\usepackage{color}
\usepackage{array}
\usepackage{multirow}
\usepackage{subfig}
\usepackage{wrapfig}
\usepackage{graphicx}
\usepackage{tabu}

\usepackage{slashed}
\usepackage{float}

\usepackage{booktabs}
\usepackage{comment}
\usepackage{enumitem}
\usepackage{graphicx}
\usepackage{siunitx}
\usepackage{xcolor}
\usepackage{verbatim}
\usepackage{dsfont}
\setlist[itemize]{leftmargin=*}

\usepackage{comment}
\usepackage{tikz}
\usetikzlibrary{decorations.pathmorphing,decorations.markings}

\tikzset{
  psi/.style={
    decoration={
      markings,
      mark=at position 0.6 with {\arrow{>}}
    },
    postaction={decorate},
    double,
    double distance=1pt
  },
  psiNoArrow/.style={
    decoration={
      markings,
      mark=at position 0.6 with 
    },
    postaction={decorate},
    double,
    double distance=1pt
  },
  nucleon/.style={
    decoration={
      markings,
      mark=at position 0.6 with {\arrow{>}}
    },
    postaction={decorate}
  },
  external/.style={},  
  gluon/.style={
  decorate, draw=black, 
  decoration={coil,amplitude=4pt, segment length=5pt}
  },
  particle/.style={draw=black, postaction={decorate}, decoration={markings,mark=at position .5 with {\arrow[draw=black]{>}}}},
 photon/.style={decorate, decoration={snake,amplitude=2pt, segment length=5pt}, draw=black}
}

\newcommand{\ComboOne}{
\begin{tikzpicture}[thick,scale=1.0]
    \draw[particle] (0,0) -- (1,0);
    \draw[particle] (1,0) -- (4,0);
    \draw[particle] (4,0) -- (5,0);
    \draw[dashed]  (1,0) arc (180:0:1.5cm) ;
    \draw[photon,green] (4.0,1.4) -- (5,2.5);

    \fill[black] (1,0) circle (0.06cm);
    \fill[black] (4,0) circle (0.06cm);

    \node at (0.5, -0.5) [external]{$\mu$};
    \node at (2.5, -0.5) [external]{$e, \mu, \tau$};
    \node at (4.5, -0.5) [external]{$e$};
    \node at (2.5, 1.8) [external]{$H_2, A$};
    \node at (4.5,2.4) [external] {$\gamma$};
\end{tikzpicture}
}

\newcommand{\ComboTwo}{%
\begin{tikzpicture}[thick,scale=1.0]
\draw[particle] (0,0) -- (1,0);
\draw[particle] (1,0) -- (4,0);
\draw[particle] (4,0) -- (5,0);
\draw[dashed] (1,0) -- (0.5+1.293,1.293);
\draw[photon,green] (4,0) -- (0.5+2.707,1.293);
\draw[particle]  (0.5+2,3)  arc (90:270:1cm);
\draw[particle]  (0.5+2,1)  arc (-90:90:1cm);
\draw[photon,green] (0.5+2,3) -- (0.5+2,4);

\fill[black] (1,0) circle (0.06cm);
\fill[black] (4,0) circle (0.06cm);
\fill[black] (2.5,3) circle (0.06cm);
\fill[black] (0.5+1.293,1.293) circle (0.06cm);
\fill[black] (0.5+2.707,1.293) circle (0.06cm);

\node at (0.5, -0.5) [external]{$\mu$};
\node at (2.5, -0.5) [external]{$e$};
\node at (4.5, -0.5) [external]{$e$};

\node at (0.76, 0.75) [external]{$H_2, A$};
\node at (3.8, 0.75) [external]{$\gamma$};
\node at (1.2,2) [external]{$t$};
\node at (3.8,2) [external]{$t$};
\node at (2.2,3.5) [external]{$\gamma$};
\end{tikzpicture}
}

\newcommand{\ComboThree}{%
\begin{tikzpicture}[thick,scale=1.0]
\draw[particle] (0,0) -- (2.5,0);
\draw[particle] (2.5,0) -- (5,0);
\draw[dashed] (2.5,0) -- (2.5,3);
\draw[particle] (0,3) -- (2.5,3);
\draw[particle] (2.5,3) -- (5,3);

\fill[black] (2.5,0) circle (0.06cm);
\fill[black] (2.5,3) circle (0.06cm);

\node at (1.25, -0.5) [external]{$N$};
\node at (3.75, -0.5) [external]{$N$};
\node at (1.25, 3.5) [external]{$\mu$};
\node at (3.75, 3.5) [external]{$e$};
\node at (2.1, 1.5) [external]{$H_2$};
\end{tikzpicture}
}

\newcommand{\be}{\begin{equation}}
\newcommand{\ee}{\end{equation}}
\newcommand{\bea}{\begin{equation}\begin{aligned}}
\newcommand{\eea}{\end{aligned}\end{equation}}

\newcommand{\hc}{\text{h.c.}}

\newcommand{\gsim}{\lower.7ex\hbox{$\;\stackrel{\textstyle>}{\sim}\;$}}
\newcommand{\lsim}{\lower.7ex\hbox{$\;\stackrel{\textstyle<}{\sim}\;$}}

\definecolor{grey}{cmyk}{0,0,0,0.75}
\definecolor{tangerine}{cmyk}{0,0.5,1,0}
\definecolor{darkgreen}{cmyk}{1,0,1,0.23}
\definecolor{Red}{rgb}{1,0,0}
\definecolor{Blue}{rgb}{0,0,1}
\definecolor{Green}{rgb}{0,1,0}
\definecolor{Grey}{cmyk}{0,0,0,0.75}
\definecolor{Tangerine}{cmyk}{0,0.5,1,0}
\definecolor{Darkgreen}{cmyk}{1,0,1,0.23}
\definecolor{Cyan}{cmyk}{1,0,0,0}
\definecolor{Yellow}{cmyk}{0,0,1,0}
\definecolor{darkblue}{cmyk}{1,0.69,0,0.11}

\newcommand{\nicpb}{Laboratory of High Energy and Computational Physics, NICPB, R\"avala pst. 10, 10143 Tallinn, Estonia}

\begin{document}

\title{Production and decays of 146~GeV flavons into $e\mu$ final state at the LHC}

\author[a]{Niko Koivunen,}
\author[a]{Martti Raidal}
%
\affiliation[a]{\nicpb} 

\emailAdd{niko.koivunen@kbfi.ee}

\abstract{
The CMS experiment at CERN has reported a possible signal for a resonance at 146~GeV decaying into the $e\mu$ final state which, presently, is the only experimental hint for lepton flavour violation in any low- and high-energy experiment. The Froggatt-Nielsen mechanism naturally predicts the existence of new scalars, the flavons, with flavour off-diagonal couplings. We study this framework in the  context of the CMS result and find that the minimal, purely leptophilic model is too restricted to match the claimed signal. Thereafter we show how models with additional flavon couplings to quarks can explain the claimed signal while satisfying all the existing constraints on lepton flavour violation.
}

\maketitle

\section{Introduction}

The CMS Collaboration has recently reported a possible signal for a new resonance at $146$~GeV decaying into charged lepton flavour violating (LFV) $e\mu$ final state\footnote{The $e\mu$ final state includes both the $e^+\mu^-$ and $e^-\mu^+$ final states.}~\cite{CMS:2023pqk,CMS:Moriond}. This is the first time when a hint for such a signal is reported, with no similar result from the ATLAS Collaboration~\cite{CMS:Moriond}, however. The claimed signal is based on  $\sqrt{s}=13$~TeV CMS data, with integrated luminosity of 138~$\rm{fb}^{-1}$. The global (local) significance of the claimed signal is $2.8\sigma$ ($3.8\sigma$) over the expected background.  At  the same time, the CMS does not find any excess in the corresponding  Higgs boson decay channel $h\to e\mu$~\cite{CMS:Moriond}.

The standard model (SM) of particle physics does not contain any new resonance nor have any sources of LFV. Therefore, if confirmed, this signal should be interpreted as a sign for new physics beyond the SM. The searches for LFV decays of $\mu$ and $\tau$ leptons, and for $\mu\leftrightarrow e$ conversion have given null result so far~\cite{Workman:2022ynf}, consistently with the SM. Similarly, no LFV decays of the SM Higgs boson have been observed.
The LFV decays $h\to e\mu$ have been constrained by the CMS and ATLAS Collaborations as $\rm{BR}(h\to e\mu)<4.4\times 10^{-5}$~\cite{CMS:2023pqk} and $\rm{BR}(h\to e\mu)<6.2\times 10^{-5}$~\cite{ATLAS:2019old}, respectively, while the CMS searches for decays $h\to e\tau$ and $h\to \mu\tau$ place upper bounds on the corresponding branching ratios as
$\rm{BR}(h\to e\tau)<2.2\times 10^{-4}$
and
$\rm{BR}(h\to \mu\tau)<1.4\times 10^{-4}$~\cite{CMS:2021rsq}. 
Consequently, if  the  claimed CMS hint  for such a new physics  will be confirmed, this would be the  first signal of LFV in any low- and high-energy experiment.

Assuming that the CMS hint for new resonance with LFV couplings actually corresponds to reality, as we do in this work, it is extremely challenging  to reconcile the claimed signal with the stringent bounds on LFV. Indeed, in this case the scale of new physics alone does not suppress any LFV process. Therefore, to explain simultaneously the presence of the claimed CMS signal and the absence of $\mu\to e\gamma$ and $\mu\leftrightarrow e$ conversion in nuclei, one must involve some additional mechanism to suppress the latter processes. For example, one could try to identify the new resonance with a spin-1 particle, like a $Z'$ with generation non-universal $U(1)$ couplings to charged leptons. This approach seems difficult, considering that there is no other degrees of freedom to cancel the $Z'$ contribution to $\mu\to e\gamma$, for example. However, in the case of complex scalars there exists a generic built-in cancellation between the scalar and pseudo-scalar components to loop-induced LFV processes. Any new physics scenario introducing LFV couplings at the scale ${\cal O}(100)$~GeV scale must involve such a mechanism to comply with observations. Therefore, we choose to work with scalars to address the new CMS hint for new physics.

The Froggatt-Nielsen (FN) mechanism~\cite{Froggatt:1978nt} is one of the most well-known methods of explaining the observed hierarchy in masses of charged fermions.  An integral part of the mechanism is the  existence of \emph{flavons}, scalars with flavour violating interactions to fermions, whose vacuum expectation values  (VEVs) generate the observed charged fermion masses and mixing via higher-order operators. The physics motivation for the Froggatt-Nielsen mechanism goes beyond collider physics. Nevertheless, it is interesting to ask whether the existence of low-scale flavons is compatible with the present particle physics phenomenology.

 In this work we study the possibility that the claimed CMS signal represents the very first experimental hint for the Froggatt-Nielsen flavon.
 First we shall concentrate on the \emph{leptophilic} flavon which only generates the flavour structure of the lepton sector and, therefore, does not couple to quarks. This allows us to avoid quite stringent additional constraints from the flavour violation in the quark sector, that would force the flavon VEV to higher scales and suppress the effects of the $e\mu$-coupling. In this set-up the SM Higgs boson and the flavon mix due to portal coupling in the scalar potential, producing two mass eigenstates, $H_1$, identified as the $125$~GeV Higgs boson and $H_2$, identified as the $146$~GeV particle. In the leptophilic Froggatt-Nielsen framework the state $H_2$, with the dominant flavon component, obtains couplings to gauge bosons and quarks through the mixing with the SM Higgs, thus allowing it to be produced at the LHC through the same processes as the SM $125$~GeV Higgs boson. Importantly, it is well established~\cite{Huitu:2016pwk} that this scenario does not have any problem to satisfy all the LFV constraints  as the pseudo-scalar contribution naturally cancels the  scalar contribution to $\mu\to e\gamma$.
  
We find that, for the maximally allowed Higgs boson-flavon mixing, the maximally allowed production cross-section for the $146$~GeV resonance is an order of magnitude smaller than the one claimed by the CMS experiment. To increase the production  cross section, we introduce additional flavour-diagonal couplings of the flavon to quarks. This, however, re-introduces the problem of LFV because, now, the tree level contribution to $\mu\leftrightarrow e$ conversion is no longer suppressed by the mixing angle. The problem can  be solved if the flavon couplings to quarks introduce additional cancellation between the tree level amplitudes to  $\mu\leftrightarrow e$ conversion. We demonstrate that this can be arranged simultaneously for the  experiments based on $Ti$ and $Au$  nuclei.

We conclude that viable Froggatt-Nielsen scenarios with low-scale flavons can be  constructed to address the hint for new $146$~GeV resonance. However, these scenarios require non-trivial model building and  some cancellation between the model parameters. More experimental data is needed to clarify the  status of the claimed CMS hint  for LFV.

\section{The leptophilic Froggatt-Nielsen model}

The claimed CMS result~\cite{CMS:2023pqk,CMS:Moriond} hints for a new resonance at the electroweak scale. The Froggatt-Nielsen mechanism should preferably be taking place at least at that scale, in order not to suffer from heavily suppressed flavon couplings. 
This does not seem to be  possible if we apply the Froggatt-Nielsen mechanism to both quark and lepton sectors simultaneously. This is due to the tree-level flavon mediated neutral meson mixing, such as the $K^0$-$\bar K^0$ mixing. The flavon VEV would then have to be close to TeV scale to avoid those constraints~\cite{Bauer:2016rxs}, which is  an order of magnitude higher than the scale indicated by the CMS experiment.  
We, therefore, apply the Froggatt-Nielsen mechanism to charged leptons only, allowing the flavon VEV to be at the electroweak scale, as demonstrated in~\cite{Huitu:2016pwk}\footnote{Each fermion sector could, in principle, have their own flavon field, with different VEVs.}.  

The Froggatt-Nielsen mechanism has been previously studied in the context of flavour violation in  the Higgs boson decays~\cite{Dery:2013rta, Dery:2014kxa,Huitu:2016pwk} and in the perspective of the LHC phenomenology~\cite{Tsumura:2009yf, Bauer:2016rxs}.

\subsection{The Froggatt-Nielsen mechanism and LFV}
The Froggatt-Nielsen framework extends the SM with a flavour symmetry which in the simplest case is global or local $U(1)$ or a $Z_N$ symmetry. We will take the flavour symmetry to be global $U(1)$. The framework introduces, as new fields,  heavy fermion messengers and a complex scalar flavon.
The SM fermions and the new particles are also charged under the flavour symmetry. The purpose of the flavour symmetry is to forbid the SM Yukawa couplings, with the possible exception for the top quark.  The Yukawa couplings are, instead, generated from effective operators. The heavy fermion messengers connect the SM fermions to Higgs boson at tree-level. Once the heavy fermion messengers are integrated out, one is left with an effective operator, which in the case of charged leptons is
\begin{equation}\label{FN operator}
\mathcal{L}\supset c_{ij}\left(\frac{\Phi}{\Lambda}\right)^{n_{ij}}\bar{L}_{L,i} H e_{R,j}+h.c.,
\end{equation}
where $c_{ij}$ are dimensionless order-one couplings, $\Lambda$ is the mass scale of the integrated out messenger fermions, $L_{L,i}$ are the  $SU(2)_L$ lepton doublets and $e_{R,i}$ are the $SU(2)_L$ charged lepton singlets, and $H$ is the SM Higgs doublet. $\Phi$ stands for a complex scalar flavon that is decomposed into real and imaginary parts as $\Phi =1/\sqrt{2}(\phi + iA)$.  The conservation of flavour charge fixes the power $n_{ij}$ as
\begin{equation}
n_{ij}=-\frac{1}{q_{\phi}}(q_{\bar L,i}+q_{R,j}+q_{H}).
\end{equation}
The power $n_{ij}$ is the number of fermion messengers that were integrated out from the diagram that generated the corresponding effective term. The $n_{ij}$ is therefore a positive integer. 

We assume that the flavon couples only to leptons, so that the  Froggatt-Nielsen mechanism generates only the charged lepton flavour structure. The operator \eqref{FN operator} gives rise to the SM Yukawa couplings, as the flavon acquires a non-zero VEV. Expanding the operator \eqref{FN operator} around the vacuum yields
\begin{eqnarray}
&&\mathcal{L}\supset c_{ij}\left(\frac{\Phi+\frac{v_\phi}{\sqrt{2}}}{\Lambda}\right)^{n_{ij}}\bar{L}_{L,i} (H+\langle H\rangle) e_{R,j}+h.c.\label{FN terms}\\
=&&c_{ij}\left(\frac{v_\phi}{\sqrt{2}\Lambda}\right)^{n_{ij}}\bar{e}_{L,i} e_{R,j}\frac{1}{\sqrt{2}}(h+v)
+n_{ij}c_{ij}\left(\frac{v_\phi}{\sqrt{2}\Lambda}\right)^{n_{ij}}\frac{v}{v_\phi}\bar{e}_{L,i} e_{R,j}~\Phi+h.c.+\cdots,\nonumber
\end{eqnarray}
where we have kept only the renormalizable terms. The first term in the second line  gives the SM Yukawa coupling
\begin{equation}
Y_{ij}\equiv c_{ij}\left(\frac{v_\phi}{\sqrt{2}\Lambda}\right)^{n_{ij}}.
\end{equation}
The charged lepton mass hierarchy is explained by 
assuming $\epsilon\equiv v_\phi/(\sqrt{2}\Lambda)<1$ and by assigning larger Froggatt-Nielsen charges to the lighter leptons compared to the heavier ones.  The flavour charge assignment determines the hierarchy of the Yukawa couplings. This is in contrast to the Standard Model where the hierarchy is obtained  by tuning the couplings themselves. 

The flavon also has Yukawa-like interaction to leptons as can be seen from the second line of Eq. \eqref{FN terms},
\be
\widetilde{\kappa}_{ij}\equiv  \frac{v}{v_\phi}n_{ij}Y_{ij}.
\ee
This coupling is not proportional to the  Yukawa matrix. It is not diagonalized simultaneously with the charged lepton  mass matrix and, therefore, the flavon couplings are flavour violating. 

The physical couplings are obtained by diagonalizing the Higgs Yukawa coupling matrix
\be
Y_{\rm diag} = U_L Y U_R^{\dagger}.
\ee
The equation \eqref{FN terms} then becomes
\be
\mathcal{L}=\frac{1}{\sqrt{2}}\bar{e}_{L} Y_{\rm diag} e_{R}(h+v)
+\bar{e}_{L} \kappa e_{R}~\Phi+h.c.,
\ee
where 
\be
\kappa_{ij} = \frac{v}{v_\phi}U_L(n\cdot Y)U_R^{\dagger},\quad \textrm{with}\quad 
(n\cdot Y)_{ij}=n_{ij} Y_{ij}.
\ee
The flavon coupling $\kappa$ can be written in a form
\be\label{kappa formula}
\kappa_{ij} = \frac{v}{v_\phi}\sum_k\left[
 Y_j q_{\bar L,k}(U_L)_{ik}(U_L^{\dagger})_{kj}
+ Y_i q_{\bar R,k}(U_R)_{ik}(U_R^{\dagger})_{kj}\right], 
\ee
where $Y_i$ is the lepton SM Yukawa coupling. 
From this expression one can deduce the maximal values for the flavour violating couplings. These  are obtained when either the left-handed or the right-handed rotation matrix  has "maximal" mixing, that is two elements $\sim 1/\sqrt{2}$ on the same row or column\footnote{This can happen if all the left- or the right-handed flavour charges are similar. If all the left-handed charges are identical, each mass matrix entry in a column has the same order of magnitude, and if all the right-handed charges are identical, each entry in each row has same order of magnitude. Even though identical charges for a handedness produces large entries in mixing matrix, the effect is lost due to unitarity. One has to break the degeneracy slightly, in order to preserve the contribution of large mixing in \eqref{kappa formula}.}. 
The maximum flavour violating flavon coupling to electron and muon therefore is
\be\label{emu max}
|\kappa^{\rm max}_{e\mu~\rm{or}~\mu e}|\sim \frac{v}{v_\phi}\frac{Y_\mu}{2}\approx 3.0\times 10^{-4}\frac{v}{v_\phi}.
\ee
This coupling is boosted if the flavon VEV is small compared to that of the SM Higgs. We will use this maximally large $e\mu$ coupling to estimate the maximum cross section obtainable in the leptophilic case in addition numerical benchmark. We set $\epsilon=0.1$ and use in both cases the flavour charges presented in Table \ref{flavour charges}.
\begin{table}[t]
\begin{center}
\begin{tabular}{|c|c|c|c|c|c|c|c|c|}
\hline
Particle   &  $e_L^c$ & $e_R$ & $\mu_L^c$ & $\mu_R$ & $\tau_L^c$ &$\tau_R$ & $H$ & $\phi$
\\
\hline 
Charge & 6 & 1 & 4 & 0 & 2 & 0 & 0 & -1\\
\hline
\end{tabular}
\end{center}
\vspace{-3mm}
\caption{The $U(1)$ flavour charges used in the  numerical and analytical estimates.}
\label{flavour charges}
\end{table}
We have chosen right-handed charges so that they are almost the same and broken the degeneracy in the first generation. This will produce large mixing  in the right-handed rotation matrix. The left-handed charges are different and produce the required hierarchy in the eigenvalues.

In the analytical estimate for the largest possible $e\mu$-cross section we assume that only off-diagonal coupling of the flavon is $\kappa_{\mu e}$ and that it is given by \eqref{emu max}. This will maximize the brancing ratio into $e\mu$. The diagonal flavon couplings in this case are approximately $\widetilde{\kappa}_{ee}\approx 6Y_e,\, \widetilde{\kappa}_{\mu\mu}\approx 4Y_\mu$ and $\widetilde{\kappa}_{\tau\tau}\approx 2Y_\tau$. We will use these values in our analytical estimate.

We also provide the following numerical benchmark that produces the correct masses for electron, muon and tau:
\be
Y=\left(\begin{array}{ccc}
4.1 \epsilon^7 & 0.63\epsilon^6 & 4.3\epsilon^6\\
6.3\epsilon^5 & -5.5927\epsilon^4 & 6.0\epsilon^4\\
-5.7\epsilon^3 & -0.2\epsilon^2 & 0.8219\epsilon^2
\end{array}\right),\quad
\widetilde{\kappa}\approx
\left(\begin{array}{ccc}
1.9\times 10^{-5} & -2.7\times 10^{-6} & 1.0\times 10^{-5}\\
-2.1\times 10^{-4} & 2.6\times 10^{-3} & -9.1\times 10^{-4}\\
3.1\times 10^{-3} & 3.5\times 10^{-3} & 2.4\times 10^{-2}
\end{array}\right).
\ee
The diagonalization matrices are given by
\be
U_L\approx\left(\begin{array}{ccc}
1 & -1.4\times 10^{-3} & -2.3\times 10^{-4}\\
-1.4\times 10^{-3} & 1 & 5.5\times 10^{-2}\\
3.0\times 10^{-4} & 5.5\times 10^{-2} & 1
\end{array}\right)\quad \rm{and}\quad
U_R\approx\left(\begin{array}{ccc}
0.55 & 0.64 & 0.54\\
-0.62 & 0.74 & -0.25\\
-0.56 & -0.20 & 0.81
\end{array}\right).
\ee

Note that the element $\widetilde{\kappa}_{\mu e}$ is $\sim 2/3$ of the maximum coupling and allows to  produce large $\rm{BR}(H_2\to e\mu)$. The flavon VEV should be $\gtrsim 100$ GeV, in order to have large enough messenger scale, $\Lambda\gtrsim 1$ TeV.

\subsection{Phenomenology at hadron colliders}

The scalar potential we work with is of the Higgs-portal type,
\be
V=
-\mu_h^2 (H^\dagger H)
-\mu_\phi^2 (\Phi^\ast\Phi)
+\lambda_h (H^\dagger H)^2
+\lambda_h (\Phi^\ast\Phi)^2
+\lambda_{h\phi}(H^\dagger H)(\Phi^\ast\Phi)
+{\mu'_\phi}^2(\Phi^2+\Phi^{\ast 2}).
\ee 
The last term in the potential explicitly breaks the $U(1)$ flavour symmetry and gives a mass, $m_A^2 = {\mu'_\phi}^2$, to the pseudo-scalar $A$, that would otherwise be a massless Goldstone boson. 

The real part of the flavon mixes with the  Higgs boson, as both the SM Higgs and the flavon acquire non-zero VEVs,
\be
\left(\begin{array}{c}
h\\
\phi
\end{array}\right)
=
\left(\begin{array}{cc}
\cos\theta & \sin\theta\\
-\sin\theta & \cos\theta
\end{array}\right)
\left(\begin{array}{c}
H_1\\
H_2
\end{array}\right).
\ee
We identify  $H_1$ as the $125$ GeV Higgs boson and  $H_2$ as the flavon like scalar. We interpret  $H_2$ as the particle with mass $146$ GeV that is responsible for the CMS signal. To constrain the mixing angle we use the  Higgs signal strength value  $\mu =1.02\pm^{0.07}_{0.06}$, obtained for $137$ $\rm{fb}^{-1}$ of  $\sqrt{s}=13$~TeV LHC data~\cite{CMS:2020gsy}.
This is compatible with the  constraint $|\sin\theta|\lesssim 0.3$.

\begin{figure}[t]
    \centering
    \includegraphics[width=0.9\textwidth]{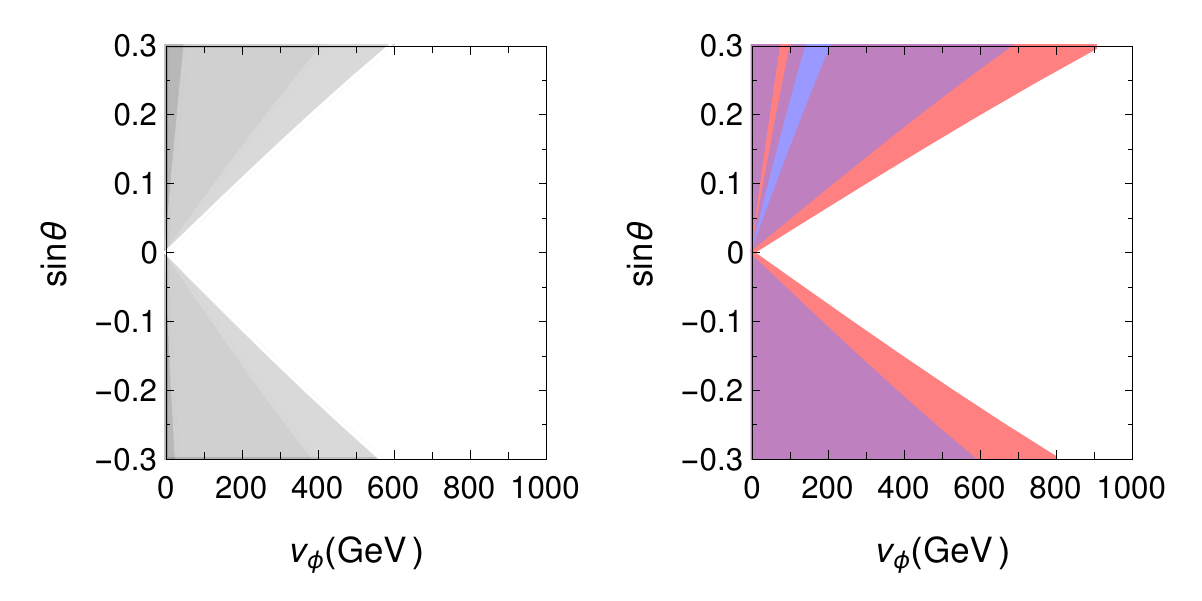}
    \caption{\emph{Left panel:} The parameter space excluded by searches for the LFV decays of Higgs boson. The dark gray area is excluded by the  $h\to e\mu$  searches~\cite{CMS:2023pqk}, the medium gray area  is excluded by the $h\to e\tau$   searches~\cite{CMS:2021rsq} and the  light gray area  by the  $h\to \mu\tau$ searches~\cite{CMS:2021rsq}. 
    \emph{Right panel:} The parameter regions excluded by the measurements of  $\rm{BR}(h\to \mu\mu)$ (blue area) and $\rm{BR}(h\to\tau\tau)$  (red  area). }
    \label{mixing vev exclusion plot}
\end{figure}

Taking into account the mixing with the Higgs boson, the lepton couplings become
\be
\mathcal{L}=\frac{1}{\sqrt{2}}\bar e_{L}\Big(\cos\theta Y_{\rm diag}-\sin\theta\kappa\Big)e_R H_1
+
\frac{1}{\sqrt{2}}\bar e_{L}\Big(\sin\theta Y_{\rm diag}+\cos\theta\kappa\Big)e_R H_2
+
\frac{i}{\sqrt{2}}\bar e_L\kappa e_R A +\hc.
\ee
The decay rate of $H_2$ to $e\mu$-final state at tree-level is
\begin{eqnarray}
\Gamma(H_2\to e\mu) = \frac{m_{H_2}}{16\pi}\cos^2\theta\left(|\kappa_{e\mu}|^2+|\kappa_{\mu e}|^2\right).
\end{eqnarray}
The mixing with flavon introduces flavor violating couplings to the $125$~GeV Higgs boson and also changes the decay rates to flavor conserving final states,
\begin{eqnarray}
\Gamma(H_1\to e_i e_i) = \frac{m_{H_1}}{16\pi}\left|\cos\theta Y_{\rm diag}^i-\sin\theta\kappa_{ii}\right|^2,\\
\Gamma(H_1\to e_i e_j) = \frac{m_{H_1}}{16\pi}\sin^2\theta\left(|\kappa_{ij}|^2+|\kappa_{ji}|^2\right).
\end{eqnarray}
These will place significant constraints on the Higgs-flavon mixing angle and on  the flavon VEV. We find that for the numerical benchmark we use the measurements of  $\rm{BR}(h\to \mu\mu)$ \cite{CMS:2020xwi} and $\rm{BR}(h\to\tau\tau)$ \cite{CMS:2017zyp} are more constraining than the searches for the LFV decays $h\to e\mu, e\tau, \mu\tau$~\cite{CMS:2021rsq, CMS:2023pqk}. The excluded parameter space in $(v_\phi,\sin\theta)$ plane is presented in Fig.~\ref{mixing vev exclusion plot}. The $h\to\mu\mu$ measurement imposes the more stringent bound compared to the one of $h\to\tau\tau$. This is due to the flavon coupling dependence on the flavour charges: the muon coupling is enhanced by larger flavour charges compared to the one of tau  lepton.  

While the flavon $\phi$ does not couple to quarks, the flavon-like mass eigenstate $H_2$ obtains coupling to quarks through mixing with the SM Higgs boson. The couplings of $H_2$ to quarks and gauge bosons are proportional to the SM couplings but scaled with $\sin\theta$. Therefore, $H_2$ can be produced at the LHC analogously to the Higgs boson, mainly in gluon-gluon fusion and vector boson fusion, which were the production channels considered in the CMS analysis.
The production cross section of $H_2$ is suppressed by $\sin^2\theta$ compared to a hypothetical SM Higgs-like scalar of a  mass $m=146$ GeV,
\be
\sigma(pp\to H_2) = \sin^2\theta ~ \sigma^{\rm SM}_{pp\to h}(m_h = 146 \rm{GeV}),
\ee
where we assume that the gluon-gluon and vector boson fusions are the only production channels.
In the narrow width approximation the production cross section for the $e\mu$ final state through $H_2$ decay is given by
\be
\sigma(pp\to H_2\to e\mu) = \sigma(pp\to H_2)\times \rm{BR}(H_2\to e\mu),
\ee
where the branching ratio is
\be\label{BR emu}
\rm{BR}(H_2\to e\mu) = \frac{\Gamma(H_2\to e\mu)}{\Gamma_{\rm tot}(H_2)},
\ee
with 
\be
\Gamma_{\rm tot}(H_2) = \sin^2\theta \big(\Gamma^{\rm SM}_{h\to \rm{all}}-\sum_{i}\Gamma^{\rm SM}_{h\to e_i e_i}\big)\Big|_{m_h=146 \rm{GeV}}+\sum_i\Gamma(H_2\to e_i e_i) + \sum_{i\neq j}\Gamma(H_2\to e_i e_j).
\ee
Here we have assumed that, $m_{H_2}<2m_A$, to prevent $H_2$ decays into imaginary part of the flavon and reducing the relevant branching ratio \eqref{BR emu}.

\begin{figure}[t]
    \centering
    \includegraphics[width=0.99\textwidth]{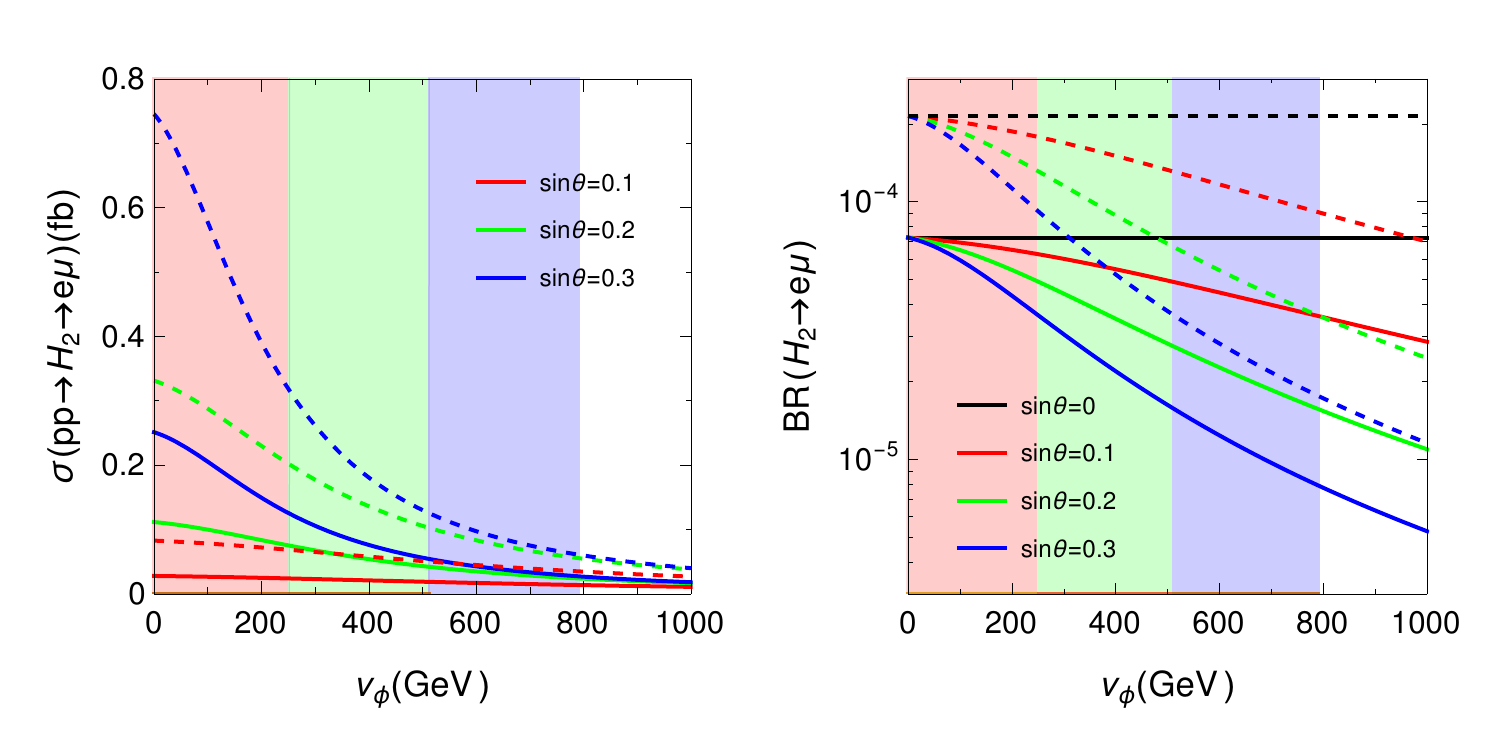}
    \caption{\emph{Left panel:} Cross-section $\sigma(pp\to H_2\to e\mu)$ as  a function of flavon VEV at 13~TeV LHC for different Higgs-flavon mixing  angles.  \emph{Right panel:}  $H_2\to e\mu$ branching ratio as  a function of flavon VEV. In both panels solid lines correspond to numerical benchmark and the dashed lines correspond to analytical estimate with maximal $e\mu$ coupling. In both panels the red area is excluded by $\rm{BR}(h\to\tau\tau)$ and $\rm{BR}(h\to\mu\mu)$ measurements for $\sin\theta=0.1$. For $\sin\theta=0.2$ red and green areas are excluded, and for $\sin\theta=0.3$ all colored areas are excluded.}
    \label{Leptophilic plot}
\end{figure}

We compute the $e\mu$ production cross section for the numerical benchmark assuming the  maximally allowed $e\mu$ coupling.
The $e\mu$ production cross section and branching ratios are presented in Fig.~\ref{Leptophilic plot}.
The gluon-gluon fusion is the dominant $H_2$ production channel, similar to the  SM Higgs production. 
The production of $H_2$ is increased with increasing the mixing angle, whereas the $\rm{BR}(H_2\to e\mu)$ is decreased when $H_2$ decays into $WW$ and $b\bar b$ final  states become more significant.
$\rm{BR}(H_2\to e\mu)$ decreases as the $v_\phi$ grows, as the leptonic couplings become smaller and the other decay channels start to dominate. 
For relatively low values of $v_\phi$, below $400$~GeV, the $\tau \bar\tau$ final state  is the dominant decay channel. This decay channel is not suppressed by $\sin\theta$, due to direct flavon coupling to leptons, and, therefore, the total $e\mu$ cross section grows with the mixing angle.

The combination of large mixing angle and small flavon VEV is excluded by the LHC data, as can be seen in Fig.~\ref{mixing vev exclusion plot}. This means that the 146~GeV particle production cross section can maximally be $\sim 0.04$~fb, over two orders of magnitude smaller  than the cross section reported by the  CMS experiment, $5.77$~fb. 
Both scenarios are in addition constrained by the searches for LFV decays of charged leptons and for $\mu\leftrightarrow e$ conversion in nuclei. These LFV processes depend on the imaginary part of the flavon, unlike the collider processes. This allows for a freedom in the parameters space to avoid those LFV constraints. These constraints are discussed in more detail in Section~\ref{sec:LFV bounds}. The parameter space compatible with these scenarios is presented in Fig.~\ref{LFV bounds} and will be discussed shortly.


\section{Addition of diagonal quark couplings}
The scenario based on leptophilic version of the Froggatt-Nielsen mechanism  is not able to produce large enough $pp\to H_2\to e\mu$ cross section, as found in the previous section. 
In the leptophilic case the flavon-like state, $H_2$, couples to quarks and gauge bosons only due to the mixing with SM Higgs, which suppresses its production.  
The $\sqrt{s}=13$~TeV LHC production cross section of the  $146$~GeV scalar with the SM couplings is $\sim 38$~pb, of which the dominant part is coming from the gluon-gluon fusion. 
In the numerical benchmark of the previous section  the branching ratio for $H_2\to e\mu$ is maximally $7.3\times 10^{-5}$. 
With this value, the production cross section of $H_2$ would have to be $\sim 80$ pb,  
in order to reach the reported  total cross section $5.77$~fb.
The flavon would need to have significant quark couplings in order to reach the  required production cross section for $H_2$. 
We will now consider adding direct quark couplings to the flavon, thus departing from the leptophilic case in order to study the Froggatt-Nielsen scenario more broadly. 
We consider the addition of flavour diagonal couplings of flavon to quarks, 
\be\label{extra terms}
\mathcal{L}_{\rm extra}=
\sum_{i=u,d,s,c,b,t}\frac{c_i Y_i}{\sqrt{2}}\bar q_{L,i} q_{R,i}\Phi+\hc,
\ee
where $Y_i$ are the SM Yukawa couplings and $c_i$ are free parameters. 

We remain agnostic about the origin of these couplings. They might originate from the Froggatt-Nielsen messenger sector in some fashion, but cannot originate from the usual Froggatt-Nielsen mechanism, analogously to \eqref{FN operator} without flavour violating couplings accompanying them. Also, the Froggatt-Nielsen mechanism would greatly limit the relative magnitude of these diagonal couplings\footnote{The application of Froggatt-Nielsen mechanism to quarks with the same flavon is out of the question due to quark flavour constraints, as  already stated above. The other possible option would be to apply the Froggatt-Nielsen mechanims to quarks by adding a second flavon field for them. In this case the leptonic flavon could have low VEV, as in the previous section, but the quarky flavon would have to have its VEV at TeV scale to avoid quark flavour constraints. The two flavons would them mix. Even if one assumes that the two flavons mix strongly, it  does not help to boost the $H_2$ production, as the quarky flavon VEV would be too suppressed by its large VEV.}. Here we simply assume that the charged leptons acquire their masses from the Froggatt-Nielsen mechanism, due to flavon $\phi$, and the quarks obtain their masses in some other fashion, perhaps via other flavon or flavons.  One way to justify the different treatment of quarks and leptons is the apparent disparity in the PMNS and CKM matrices, the former exhibiting order one elements and the latter hierarchical elements. Also the lepton sector shows more drastic differences in masses, neutrinos being at least six orders of magnitude lighter than electron. Nevertheless, we assume that the mass generation of charged leptons and quarks are linked to each other in some fashion and, hence, the couplings in \eqref{extra terms}  arise.

Now that we have introduced direct couplings to quarks for the flavon, we set the mixing angle with  the SM Higgs boson to zero as it is no longer required for $H_2$ production at the LHC. From now on the mass eigenstate $H_2$ is the pure flavon. As there is no mixing, the properties of the  Higgs boson stay those of the SM and we are free from constraints presented in Fig.~\ref{mixing vev exclusion plot}.
One might consider the production of $H_2$  directly from the light quarks, instead of gluon-gluon fusion through the top-loop. Relatively large couplings to up and down, $c_u Y_u \sim c_d Y_d\sim 5\times 10^{-2}$, would yield the desired $H_2$ production cross section $\sim 80$~pb. The large up and down couplings will also boost those $H_2$ decay channels that suppress the $e\mu$ branching ratio, effectively killing this signature. The light quarks cannot, therefore, be used for producing the flavon.   

The $H_2$ production cross section needs to be increased at least by the factor of $2$, compared to $146$ GeV SM Higgs-like scalar, in order to reach the cross section indicated by  the CMS experiment. 
This can be accomplished with flavon coupling to top quark of the magnitude $\sim \sqrt{2}$. This coupling alone would ensure sufficient production cross section for $H_2$. The flavon cannot decay directly into top-antitop pairs, thus avoiding the suppression of $H_2\to e\mu$ branching ratio. The large top coupling, however, makes the $H_2\to gg$ decay rate significant for large values of $v_\phi$. 
Couplings other than to the top quark are not required for efficient production of $H_2$. However, they are required to avoid  flavour constraints that become more restrictive due to the additional quark-flavon couplings.


\subsection{Constraints from the  LFV searches}
\label{sec:LFV bounds}

\begin{figure}[t]
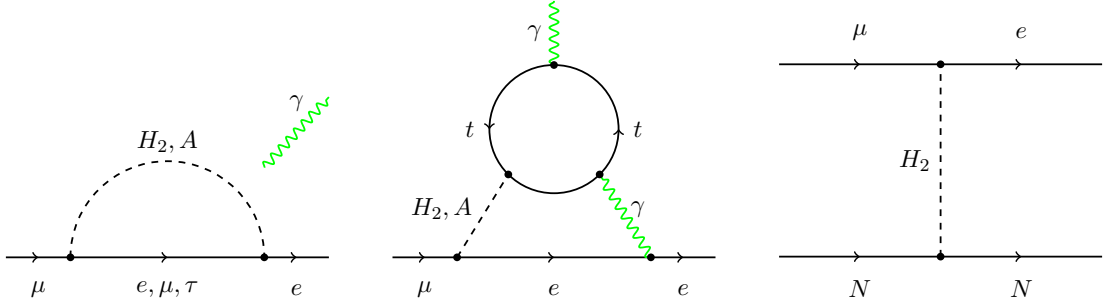

    \centering
    \scalebox{0.85}{\ComboOne 
    \hspace{0.7cm}
    \ComboTwo
    \hspace{0.7cm}
    \ComboThree}
    \caption{\emph{Left and middle panels:} The $1$- and $2$-loop contributions to $\mu\to e\gamma$ in the case of additional quark couplings. \emph{Right panel:} The tree-level contribution to $\mu\leftrightarrow e$ conversion in nuclei  in the case of additional quark couplings.}
    \label{LFV diagrams}
\vspace*{-2mm}
\end{figure}

The constraints from LFV decays of leptons are the most stringent in the $(e,\mu)$ sector, unlike in the case of SM Higgs for which they are the weakest (as seen in Fig.~\ref{mixing vev exclusion plot}).
The searches for LFV decays of muon $\mu\to e\gamma$~\cite{MEG:2016leq}, $\mu\to eee$~\cite{SINDRUM:1987nra} and various tau decay channels 
impose stringent constraints on the LFV couplings. 
In addition, the $\mu\leftrightarrow e$ conversion in various nuclei, such as gold \cite{SINDRUMII:2006dvw} and titanium \cite{SINDRUMII:1993gxf}, impose relevant constraints. In our analysis of LFV decays and $\mu\leftrightarrow$ conversion we follow references \cite{Huitu:2016pwk, Harnik:2012pb, Kitano:2002mt}.
We find that the $\mu\to e\gamma$ and $\mu\leftrightarrow e$ conversion in gold impose the most stringent constraints on the scenarios we study. 

In this  scenario,  $\mu\to e\gamma$ acquires significant and comparable contributions  both at 1-loop and 2-loop level. In the leptophilic model with Higgs-flavon mixing the contributing diagrams along with the associated formulae are presented in Ref.~\cite{Huitu:2016pwk}. Here  we shall concentrate on the case with additional flavon-quark couplings. The relevant diagrams in the case of additional quark couplings are presented in the left and in the middle panels of Fig.~\ref{LFV diagrams}.  The 2-loop Barr-Zee diagram \cite{Barr:1990vd} contains large top coupling that compensates for the additional loop suppression. Both $1$- and $2$-loop diagrams are mediated by real and imaginary parts of the flavon. Their contributions come with opposite signs, yielding cancellations in certain regions of the parameter space, allowing to bypass the stringent $\mu\to e\gamma$ constraint, despite of large flavour violating coupling and relatively low mediator masses.

\begin{figure}[t]
    \centering
    \includegraphics[width=0.95\textwidth]{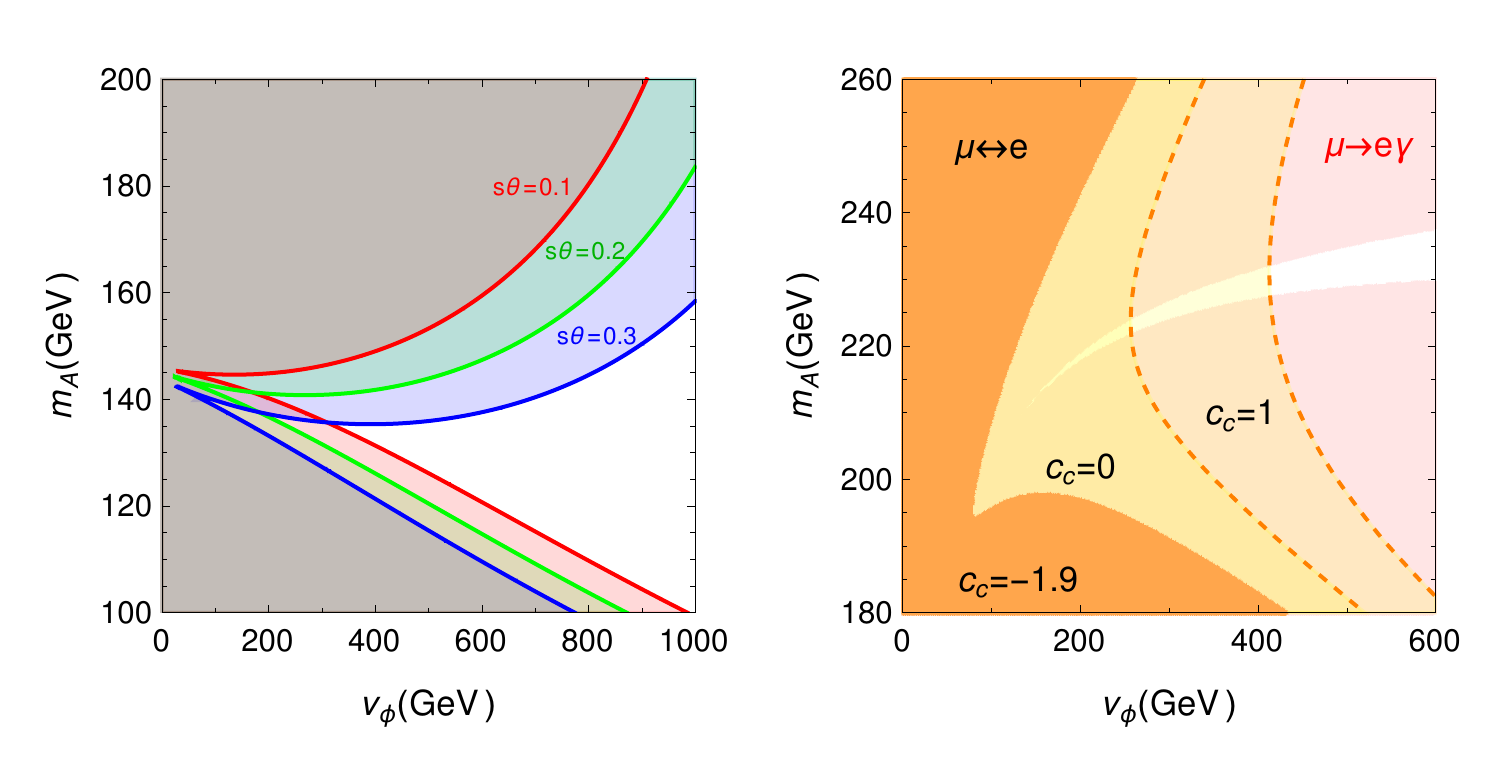}
    \caption{\emph{Left panel}: Regions of the $(v_\phi,\,m_A)$ parameter  space excluded by searches for $\mu\to e\gamma$  in the  leptophilic model  for different values of the mixing angle. \emph{Right panel}: The LFV constraints in the case of additional quark-flavon couplings for $c_t= 1.9$ and three different values of the charm coupling, $c_c = -1.9$, $0$ and $1$. The allowed white wedge shaped area continues until the corner of the orange area, but is too fine to be visible by eye.}
    \label{LFV bounds}
\end{figure}

The $\mu\to e\gamma$ decay rate is  given by
\be
\Gamma(\mu\to e\gamma)=\frac{m_\mu^3}{4\pi}(|A_L|^2+|A_R|^2),
\ee
where
\be\label{dipole contribution}
A_{L,R}=A_{L,R}^{1-\rm{loop}}+A_{L,R}^{2-\rm{loop}}.
\ee
In the case of additional quark couplings the $1$- and $2$-loop contributions are
\bea\label{1-loop contribution}
A_L^{1-\rm{loop}}=
& \sum_{i=e,\mu,\tau}\left(-\frac{ie}{64\pi}\right)\kappa_{ie}^\ast\int_0^1\frac{\kappa_{i\mu}y(y-x)m_\mu+\kappa_{\mu i}^\ast(y-1)m_i}{y(y-x)m_\mu^2+(1-y)m_i^2+m_{H_2}^2y},\\
+
& \sum_{i=e,\mu,\tau}\left(\frac{ie}{64\pi}\right)\kappa_{ie}^\ast\int_0^1\frac{-\kappa_{i\mu}y(y-x)m_\mu+\kappa_{\mu i}^\ast(y-1)m_i}{y(y-x)m_\mu^2+(1-y)m_i^2+m_A^2y},
\eea
and
\bea\label{2-loop contribution}
A_L^{2-\rm{loop}}=
-i\frac{e\alpha G_F v}{12\pi^3}\kappa_{\mu e}^\ast c_t \Big[f(z_{t\phi})-g(z_{tA})\Big],
\eea
with 
\bea
&f(z)=\frac{z}{2}\int_0^1\frac{1-2x(1-x)}{x(1-x)-z}\log\left[\frac{x(1-x)}{z}\right],\\
& g(z)= \frac{z}{2}\int_0^1\frac{1}{x(1-x)-z}\log\left[\frac{x(1-x)}{z}\right].
\eea
The arguments of these functions are defined by $z_{t\phi}=m_t^2/m_{H_2}^2$ and $z_{tA}=m_t^2/m_A^2$.
The $A_R$-terms are obtained from $A_L$ by replacing $\kappa_{ij}$ with $\kappa_{ji}^\ast$. Note the sign differences between $H_2$ and $A$ contributions in Eqs. \eqref{1-loop contribution} and \eqref{2-loop contribution}. As the pseudo-scalar $A$ does not enter the relevant collider processes, its mass is an independent parameter. This fact allows for the cancellations between different amplitudes and, thus,  suppression of $\mu\to e\gamma$ decays.


The $\mu\leftrightarrow e$-conversion rate is  given by
\bea
\Gamma(\mu\leftrightarrow e) &=
\left|\frac{iD}{2m_\mu}A_L
+\widetilde{g}^{(p)}_{LS} S^{(p)}
+\widetilde{g}^{(n)}_{LS} S^{(n)}
+\widetilde{g}^{(p)}_{LV} V^{(p)}
\right|^2\\
&+\left|\frac{iD}{2m_\mu}A_R
+\widetilde{g}^{(p)}_{RS} S^{(p)}
+\widetilde{g}^{(n)}_{RS} S^{(n)}
+\widetilde{g}^{(p)}_{RV} V^{(p)}
\right|^2.
\eea
The $\mu \leftrightarrow e$  coversion receives contributions generated by 1-loop and 2-loop diagrams in Fig.~\ref{LFV diagrams} and the coefficients $A_L$ and $A_R$ are given in Eqs.~\eqref{dipole contribution}, \eqref{1-loop contribution} and \eqref{2-loop contribution}. The $\mu \leftrightarrow e$ conversion receives possibly dominant tree-level contribution presented by the right diagram in Fig.~\ref{LFV diagrams}. The tree-level process does not receive contribution from pseudo-scalar $A$, as its contribution vanishes for coherent scattering~\cite{Kitano:2002mt}. Finally, there is sub-leading vector contribution with expression given in Ref.~\cite{Harnik:2012pb}. 

The contributions from tree-level scalar interactions (Fig.~\ref{LFV diagrams}) with the proton and the neutron are:
\bea\label{scalar conversion}
\widetilde{g}^{(p)}_{LS}=-\frac{\sqrt{2}}{m_{H_2}^2}\frac{m_p}{v}\kappa_{e\mu}\sum_i c_i f^{(i,p)},\quad
\widetilde{g}^{(n)}_{LS}=-\frac{\sqrt{2}}{m_{H_2}^2}\frac{m_n}{v}\kappa_{e\mu}\sum_i c_i f^{(i,n)}.
\eea
The summation is over all the quarks: $i=u,d,s,c,b,t$, and $m_p$ and $m_n$ are the proton and neutron masses respectively.  
The $\widetilde{g}^{(p)}_{RS}$ and $\widetilde{g}^{(n)}_{RS}$ are obtained from $\widetilde{g}^{(p)}_{LS}$ and $\widetilde{g}^{(n)}_{LS}$ by replacing $\kappa_{e\mu}$ with $\kappa_{\mu e}^\ast$. The overlap integrals for gold\footnote{The overlap integrals for other nuclei can be found in \cite{Kitano:2002mt}.} are $D=0.189$, $S^{(p)}=0.0614$, $S^{(n)}=0.0918$ and $V^{(p)}=0.0974$, in units of $m_\mu^{5/2}$. The nucleon matrix elements for light quarks are 
\be
f^{(u,p)}=f^{(d,n)}=0.024,\quad
f^{(d,p)}=f^{(u,n)}=0.033,\quad
f^{(s,p)}=f^{(s,n)}=0.25,
\ee
and 
\be\label{heavy matrix elements}
f^{(c,p)}=f^{(c,n)}=f^{(b,p)}=f^{(b,n)}=f^{(t,p)}=f^{(t,n)}
=
0.051,
\ee
for the heavy quarks \cite{Harnik:2012pb}.

 In the leptophilic model the quark coupling to flavon is suppressed by $\sin\theta$ and, therefore, the tree-level contribution to $\mu\leftrightarrow e$-conversion is not competitive with the dipole contribution $A_{L,R}$. 
 In the leptophilic case the $\mu\to e\gamma$ provides the most stringent constraint on the model parameters.
 The allowed parameter space in leptophilic model is presented in the left panel of Fig.~\ref{LFV bounds}. The effect of cancellation of scalar $H_2$ and pseudo-scalar $A$ is clearly visible: large flavon couplings (small $v_\phi$) are allowed for $m_A\sim m_{H_2}$, where the cancellation is most effective. 

For the case of additional quark couplings the situation is more involved.
The process $\mu\to e\gamma$ enjoys the cancellation between real and imaginary parts of flavon, just like in the leptophilic case, and the $\mu\to e\gamma$ bound can be avoided at small $v_\phi$, even for the large flavon-top coupling.  The $\mu\leftrightarrow e$ conversion, however, excludes the small flavon VEVs in this case. This can be alleviated by adding flavon coupling to quark(s), other than top, with an opposite sign. This can cancel the large top contribution to  $\mu\leftrightarrow e$ conversion.
In the light of Eqs.~\eqref{scalar conversion} and \eqref{heavy matrix elements}, the cancellation between two tree-level contributions of heavy quarks takes place when the coupling parameters $c_i$ have the same absolute value but come with opposite signs. The cancellation of top contribution is also possible with other quarks. For the flavour constraints the choice between $b$ and $c$ quark is not relevant, but for later collider results it is. The bottom SM Yukawa coupling is larger than that of charm. The addition of bottom will, therefore, dilute the $H_2\to e\mu$ branching ratio more. We will, therefore, choose charm to cancel the top contribution.

We study the flavon production through gluon-gluon fusion at the  LHC and set the top coupling to be $c_t=1.9$.
The results  are shown in the right panel of Fig.~\ref{LFV bounds}: for the opposite contributions of top and charm the tree-level contributions cancel, leaving $\mu\to e\gamma$ as the dominant process. In this case low values of $v_\phi\sim 100$ GeV are allowed. With top coupling only, the $\mu\leftrightarrow e$ conversion  constrains the low flavon VEVs, exluding VEVs below $\sim 260$~GeV. For charm coupling equal to the SM Higgs charm coupling, the flavon VEVs below $\sim 410$~GeV are excluded.  


\subsection{Collider results}

\begin{figure}[t]
    \centering
    \includegraphics[width=0.98\textwidth]{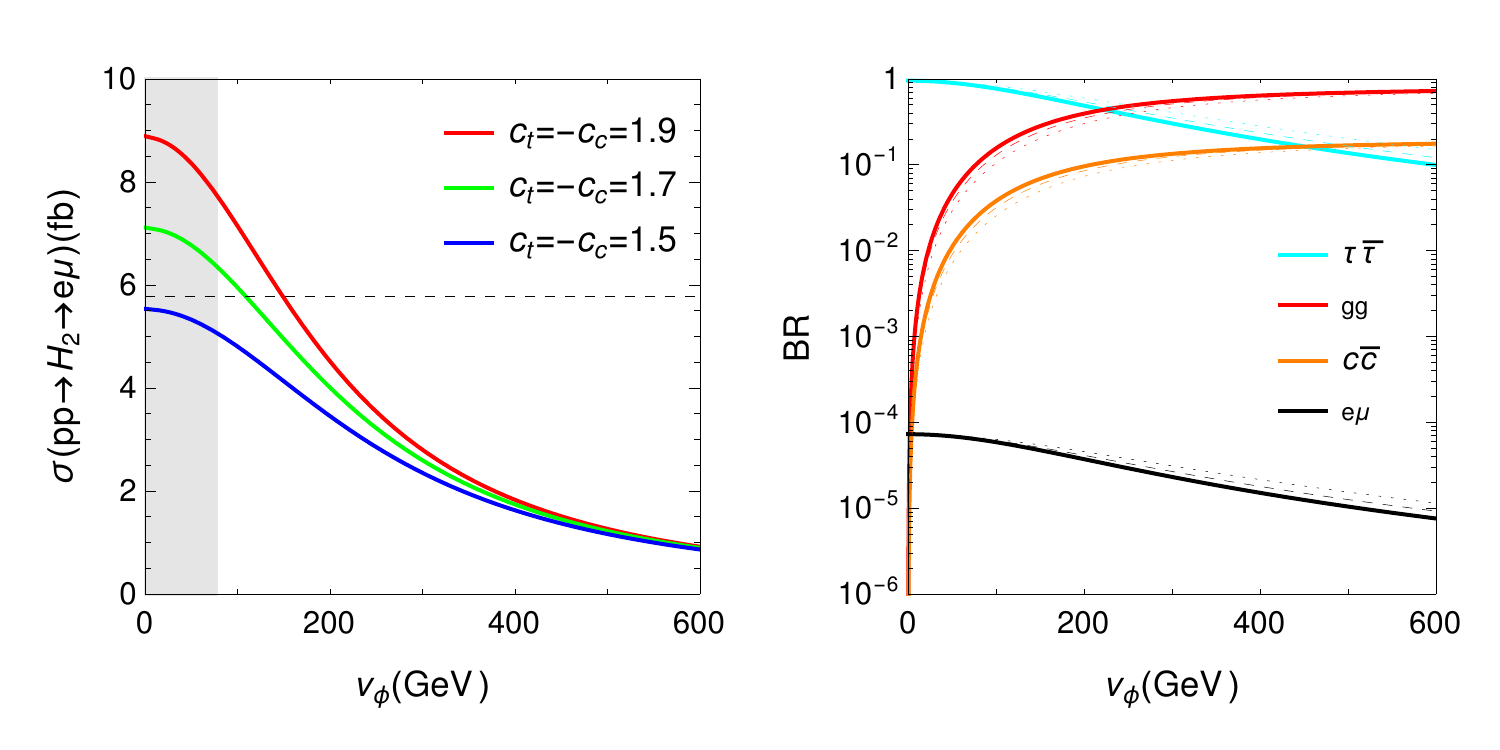}
    \caption{\emph{Left panel:} Cross-section $\sigma(pp\to H_2\to e\mu)$ as a function of the flavon VEV at the $\sqrt{s}=13$ TeV LHC for different quark couplings as indicated in the  figure. The gray shaded region is excluded by $\mu\leftrightarrow e$ conversion for all values of $m_A$. The horizontal line corresponds to $5.77$~fb. \emph{Right panel:} The relevant branching ratios for the same case. The solid lines correspond to $c_t=-c_c=1.9$, dashed line to $1.7$ and dot dashed line to $1.5$.}
    \label{quarky}
\end{figure}

We consider here the flavon production through gluon-gluon fusion with large coupling to top quark. We also have flavon coupling to charm, in order to avoid the  flavour bounds, but it offers negligible contribution to flavon production.  
As there is no mixing with the  Higgs boson, the relevant $H_2$ production channels are the dominant gluon-gluon fusion and the  sub-dominant $tt$-fusion. The resulting $e\mu$ production cross section is presented in Fig.~\ref{quarky}. One can see that large flavon coupling to top, $c_t Y_t\gtrsim 1.7$, can reproduce the reported CMS cross section for the $e\mu$ final state for $v_\phi\gtrsim 100$ GeV. 

Other  possible collider constraints on the $146$ GeV flavon are avoided because there is no mixing with the  SM Higgs boson and, therefore, $H_2$ does not couple directly to gauge bosons. Therefore the searches for $ZZ$ final state from the second Higgs boson at low mass~\cite{CMS:2018amk} do not impose constraints. 
Most other second scalar searches by the CMS and ATLAS impose limits on masses above $146$~GeV.  The ATLAS searches for $WW$ and $ZZ$ final states place a limit on masses above $300$~GeV~\cite{ATLAS:2022eap}, whereas the CMS search for $WW$ final state set a limit above $200$~GeV~\cite{CMS:2019bnu}.
Finally, the $\gamma\gamma$ final state searches restrict masses above $200$~GeV (ATLAS)~\cite{ATLAS:2021uiz} and $500$~GeV (CMS)~\cite{CMS:2018dqv}. As a result, the Froggatt-Nielsen scenario with additional quark couplings is, therefore, not constrained by other collider searches, even though the required top coupling is large.




\section{Conclusions}
\label{conclusion}

In this work we studied the phenomenology of low-energy Froggatt-Nielsen mechanism with the aim to address the  recent CMS hint for a new $146$~GeV
resonance with LFV couplings. We first studied the purely leptophilic model which is known to be consistent  with all stringent constraints on LFV. We found that the leptophilic Froggatt-Nielsen model cannot reach the production cross section $\sigma=5.77$~fb which is indicated by the CMS result. The simplest realizations of the Froggatt-Nielsen mechanism are, thus, incompatible with the claimed CMS signal. 

We modified the purely leptophilic model by adding diagonal  flavon couplings to quarks. A large flavon coupling to top quark allows for sufficient production of flavon and the experimentally hinted cross-section can be obtained. Additionally, couplings to lighter quarks are also required in order to avoid the bounds  arising from tree level mediated $\mu\leftrightarrow e$ conversion. The latter can be avoided when the lighter quark couplings to flavon have different sign compared to the top coupling.   

We conclude that non-trivial model building and some cancellation between the  model parameters are needed to identify the CMS hint for a new resonance with the low-scale Froggatt-Nielsen flavon. More experimental data is needed to clarify the origin and properties of the claimed CMS signal.

\vspace{0.5cm}

{\bf Acknowledgements.}  
This work was supported by the Estonian Research Council grants PRG803 and PRG1677. 

\vspace{0.5cm}

{\bf Note added.} The CMS hint for the 146~GeV resonance as a new scalar particle has also been studied in Ref.~\cite{Primulando:2023ugc} in the context of two Higgs doublet models.

\bibliographystyle{JHEP}
\bibliography{collider_331}

\end{document}